\documentclass[12pt]{article}
\usepackage[cp866]{inputenc} 
\usepackage[english]{babel} 

\newcommand{\be}{\begin{equation}}
\newcommand{\ee}{\end{equation}}
\newcommand{\ba}{\begin{eqnarray}}
\newcommand{\ea}{\end{eqnarray}}

\author{V. ~A.~ Petrov\\
        {\it  Institute for High Energy Physics, Protvino, RF}}
        
\title{On Vector Dominance}

\begin{document}

\maketitle

\begin{abstract}
We argue that a Regge-pole modification of the VDM puts on equal footing the asymptotic behaviour of electromagnetic
and neutral vector meson currents.
\end{abstract}

Vector Dominance Model (VDM) is based on proportionality of the operator of electromagnetic current $J_{\mu}^{em}$ and field operators of neutral 
	vector mesons $V_{\mu}$ :

\be
J_{\mu}^{em} = \sum_V c_V  V_{\mu}
\ee

           where  $c_V$ is a constant $c$-number coefficient [1].

There were a lot of uses of this formula, we don't concern them at the moment. Let's concentrate on the use of this formula for description of hadronic  electromagnetic form factors. Straightforward application of this formula to the pion form factor (neglecting the widths of vector mesons) reads :

\be
F(t) = \sum_V \frac{c_V}{m_V^2-t} \Gamma_V(t)
\ee
 
where

\ba  
\langle p'|J^{em}_\mu|p\rangle &=& (p' + p)_{\mu} F_{\pi}(t), \nonumber\\    
\langle p'|I^{V}_\mu|p\rangle &=& (p' + p)_{\mu} \Gamma_{V}(t), \nonumber\\
t &=&  (p' - p)^2,\nonumber\\
I^V_\mu &=& (m^2_V - \partial^2)V_\mu.\nonumber
\ea

From the celebrated constituent-counting rule (CCR)~~[2]~~ it follows that at high values\\ of $t$ 

\be
F_\pi (t) \sim 1/t.
\ee

Quite often, assuming $\Gamma_V (t) = const$, this behavior was "explained" by Eq.(2). Let us, however, notice that all the arguments of CCR applied to 
$\langle p'|J^{em}_\mu|p\rangle$ to derive the asymptotics (3) are to the same extent applicable to the matrix element
 $\langle p'|I^{V}_\mu|p\rangle$. Thereof one gets immediately that at  $- t  \rightarrow \infty$
 
$$
\Gamma_V (t)\sim 1/t.
$$
Thus we are faced with a contradiction, since the simultaneous fulfillment of such a behavior both for $F_\pi (t)$ and $\Gamma_V (t)$
 is impossible as implied by Eq. (2).

Let's consider the problem from another perspective. When analyzing deep inelastic processes one deals - 
via the operator product expansions - 
with composite spin-$J$ operators $O^i_{\mu_1...\mu_J}$  of the type 
$$
O_{\mu_1...\mu_J}^i \sim \vec{q} \gamma_{\mu_1}\lambda^iD_{\mu_2}...D_{\mu_J}q
$$
where $q$ stands for the quark field of flavor $q$. In particular (we limit us with "light" flavor SU(3)), 

$$
J_{\mu}^{em} = \sum_q e_q \vec{q}  \gamma_{\mu} q = O_{\mu}^3 + O_{\mu}^8 = \vec{q}\gamma_{\mu}\frac{1}{2}\lambda^3 q +
\vec{q}\gamma_{\mu}\frac{1}{2\sqrt{3}}\lambda^8 q
$$
where $e_q$ are quark electric charges in units of positron charge. I.e. the current $J_{\mu}^{em}$ is a spin-1 non-singlet composite operator.

If we take the matrix element 
$$
\langle p'|O^{i}_{\mu_1...\mu_J}|p\rangle = \lambda_{\mu_1...\mu_J}(p', p)f^q_J(t)
$$ 

then, in this terminology,
$$
F_{\pi}(t) = \sum_q e_q  f_1^q (t).
$$

If we consider the quantity $f_J^q (t)$ as a function of $J$ continued to the complex plane then  it has to possess a Regge-pole singularity:

$$
f_J^q (t) \sim \frac{1}{J-\alpha_H(t)}
$$
where $\alpha_H (t)$ is the Regge trajectory containing hadrons $H$ with quarks $q$ as valence  constituents.
In our case we have to have
$$ 
f_1^q (t)\sim \frac{1}{1-\alpha_V(t)}
$$
where $\alpha_V (t)$ runs over $\alpha_{p}(t), \alpha_{\omega}(t), \alpha_{\varphi}(t)$  trajectories.

At  $t \rightarrow m_V^2$ 

$$
\alpha_V(t) \rightarrow 1 + \alpha'_V(m_V^2)(t - m_V^2) + ...
$$
and we reproduce the standard VDM factor.

The corresponding generalization of the VDM for the pion form factor could  look as follows
$$
F_{\pi} (t) = \sum_V \frac{c_V\alpha'_V(m^2_V)}{1-\alpha_V(t)}\Gamma_V(t)
$$
Actually  equation of this kind should hold for the form factor of any hadron, not only pions.

It is worth to notice that similar generalizations of the form factor were suggested long before but either with
$\Gamma_V (t)\approx const$ [3] or with some specific $ad$ $hoc$ assumptions about $\Gamma_V (t)$ [4]. 
The general fact that form factors should contain Regge trajectories can be found also
 in some  vague form in the known monographs [5].

 Now let's see what have we gained from such a modification of the VDM.
At first sight, taking into account  the celebrated  linearity of  Regge trajectories
$$
\alpha_V (t) = 1 + \alpha'_V (t - m_V^2),
$$
almost nothing because when dealing with asymptotic behavior we encounter the same inconsistency as we had with usual VDM:
$$
\frac{1}{t} \sim \frac{1}{t}\frac{1}{t}.
$$
In compare with the VMD version with  $\Gamma_V (t) = g_{\pi\pi V}$ or  $\frac{m_V^2}{m_V^2 - t}g_{NNV}$ 
our conjecture is less informative and less falsifiable because we only relate two apriori  unknown vertices,
 $F(t)$ and  $\Gamma_V (t)$, with help of the factor  $\frac{c_V \alpha'_V(m_V^2)}{1 - \alpha_V(t)}$  which is known.

Nonetheless, things go not so bad. There are serious arguments  that meson Regge trajectories have to tend to constant values at 
$- t \rightarrow \infty$ [6]. This circumstance removes the puzzle of asymptotic incompatibility  between $F_{\pi} (t)$ and 
$\Gamma_V (t)$. This means in particular that processes induced by highly virtual photons are equivalent to processes induced by highly virtual vector mesons. Normally VMD  was deemed applicable only at not very high values of $t$. In a modified form this restriction is lifted.

This circumstance allows one to extend the unitarity implications derived for off-shell vector-meson scattering amplitudes to
	 photon-proton scattering amplitudes [7].
\vskip 5mm

I am grateful to A.A. Godizov and R.A. Ryutin for useful discussions.


\newpage

\begin{thebibliography}{99}
\bibitem{1}
\textit{J.J. Sakurai.} Currents and Mesons. The University of Chicago Press,
Chicago 1969.

 \bibitem{2}
\textit{V. A. Matveev,  R.M.Muradyan and A. N. Tavkhelidze.} Lett. Nuovo.Cim.7(1973)719.

\textit{J. Brodsky and G.R. Farrar.} Phys.Rev.Lett.31 (1973) 1153.

\bibitem{3}
\textit{M. McMillan and E. Predazzi.}  Nuovo Cimento 25(1962) 838 

\bibitem{4}
\textit{P. Di Vecchia, F. Drago.} Lett.Nuovo Cim. 1 (1969) 917.

\bibitem{5}
\textit{S. L. Adler and R.F. Dashen.} Current Algebras and Applications to Particle Physics. W.A. Benjamin Inc., New York-Amsterdam, 1968.
Ch.5, \S 3.

\textit{V. De Alfaro, S. Fubini, G. Furlan and C. Rossetti.} Currents in Hadron Physics. North-Holland Publishing Company, Amsterdam - London, 1973. Ch.4, \S 8.
\bibitem{6}
\textit{J. Kwiecinski.} Phys.Rev.D 26(1982); P.D.B. 
\textit{Collins and P.J. Kearny.} Z. Phys. C22(1984)277;
\textit{R. Kirschner and L. N. Lipatov.} Z. Phys. C45(1990);
\textit{S.J. Brodsky, W.-K. Tung and C. B. Thorn.} Phys.Lett.B318(1993)203;
\textit{R. Kirschner.} Z.Phys. C67(1995);
\textit{A. A. Godizov and V. A. Petrov.} JHEP 0707 (2007) 083.

\bibitem{7}
\textit{V.A. Petrov.} Proc.VIth Blois Workshop. Gif-sur-Yvette, Ed.       Fronti\`eres, 1996. P.139;\\
\textit{A. V. Prokudin.} Proc.VIIIth Blois Workshop. World Scientific Publishing      Company. 2000. P.95.


\end{thebibliography}
\end{document}